\documentclass[aps, prl, 
amsmath, showpacs, preprintnumbers, twocolumn, amssymb, amsmath, superscriptaddress]{revtex4-1}
\usepackage{graphicx}
\usepackage{mathptmx, textcomp}
\usepackage[latin1]{inputenc}
\usepackage{tabularx}
\usepackage{units}
\usepackage{color}
\usepackage{bm}

\begin{document}
\title{Recoil-limited feedback cooling of single nanoparticles near the ground state in an optical lattice}
\author{M.\,Kamba}
\affiliation{Department of Physics, Tokyo Institute of Technology, Ookayama 2-12-1, Meguro-ku, 152-8550 Tokyo}
\author{H.\,Kiuchi}
\affiliation{Department of Physics, Tokyo Institute of Technology, Ookayama 2-12-1, Meguro-ku, 152-8550 Tokyo}
\author{T.\,Yotsuya}
\affiliation{Department of Physics, Tokyo Institute of Technology, Ookayama 2-12-1, Meguro-ku, 152-8550 Tokyo}
\author{K.\,Aikawa}
\affiliation{Department of Physics, Tokyo Institute of Technology, Ookayama 2-12-1, Meguro-ku, 152-8550 Tokyo}

\date{\today}


\begin{abstract}
We report on direct feedback cooling of single nanoparticles in an optical lattice to near their motional ground state. We find that the laser phase noise triggers severe heating of nanoparticles' motion along the optical lattice. When the laser phase noise is decreased by orders of magnitude, the heating rate is reduced and accordingly the occupation number is lowered to about three. 
We establish a model directly connecting  the heating rate and the measured laser phase noise and elucidates that the occupation number under the lowest laser phase noise in our system is limited only by photon recoil heating. Our results show that the reduction of the laser phase noise near the oscillation frequency of nanoparticles is crucial for bringing them near the ground state and pave the way to sensitive accelerometers and quantum mechanical experiments with ultracold nanoparticles in an optical lattice.
\end{abstract}

\maketitle



Decelerating the motion of microscopic particles such as atoms and molecules has been an essential ingredient for revealing their fundamental properties and for realizing their applications in various fields, ranging from precision measurements~\cite{derevianko2011colloquium,ludlow2015optical} and fundamental physics~\cite{safronova2018search} to quantum information processing~\cite{haffner2008quantum,saffman2010quantum} and quantum simulations~\cite{bloch2008many,dalibard2011colloquium}. In comparison with microscopic particles, decelerating the motion of mesoscopic and macroscopic particles has been a challenging task because of the lack of the efficient mechanism for deceleration.  Nanometer-sized objects prepared near their motional ground state are expected to be a promising system for various applications including force sensing~\cite{ranjit2016zeptonewton,hempston2017force,hebestreit2018sensing}, the test of superposition states at macroscopic scales~\cite{romero-isart2011large,bassi2013models}, detecting gravitational waves at high frequencies~\cite{arvanitaki2013detecting}, and the search for non-Newtonian gravity forces~\cite{geraci2010short}.

In recent years, remarkable progresses have been made in decelerating the motion of nanoparticles in vacuum either by feedback cooling or by cavity cooling. In the former scheme, a time-varying force opposite to the velocity of particles is applied~\cite{li2011millikelvin,gieseler2012subkelvin,vovrosh2017parametric,iwasaki2019electric,tebbenjohanns2019cold,conangla2019optimal}, whereas the latter scheme relies on photon scattering in a high-finesse cavity transferring the motional energy of particles to photons ~\cite{kiesel2013cavity,asenbaum2013cavity,millen2015cavity,meyer2019resolved}. Very recently, these approaches have been applied for cooling nanoparticles in a single-beam optical trap to near their motional ground state~\cite{delic2020cooling,tebbenjohanns2020motional}. Aside from a single-beam optical trap, an optical lattice, a standing wave potential created by retro-reflecting a laser, has been a crucial tool for trapping and manipulating atoms and molecules~\cite{bloch2008many,dalibard2011colloquium}. However, in spite of its possibility in diverse quantum mechanical experiments thanks to the high oscillation frequency in an optical lattice, the use of an optical lattice for nanoparticles has been limited~\cite{ranjit2016zeptonewton,yoneda2018spontaneous,iwasaki2019electric}.

\begin{figure}[t]
\includegraphics[width=0.9\columnwidth] {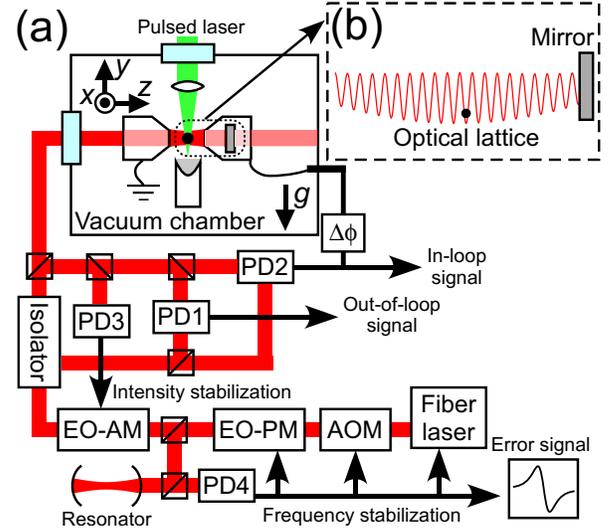}
\caption{(color online).  (a) Schematic representation of our experimental setup. Single nanoparticles are trapped in an optical lattice formed by a single-frequency fiber laser at $\lambda=\unit[1550]{nm}$, which is intensity-stabilized by an electro-optic amplitude modulator (EO-AM) and frequency-stabilized by an electro-optic phase modulator (EO-PM) and an acousto-optic modulator (AOM). The motion of nanoparticles is observed via photodetectors (PDs). (b) Blow up near the trapping region. Nanoparticles are located at the focus of the incident beam where the oscillation frequency along the optical lattice is the highest. The position fluctuation of nanoparticles due to fluctuations in the laser frequency scales linearly with the distance from the retro-reflecting mirror. }
\label{fig:expset}
\end{figure}

\begin{figure}[t]
\includegraphics[width=0.95\columnwidth] {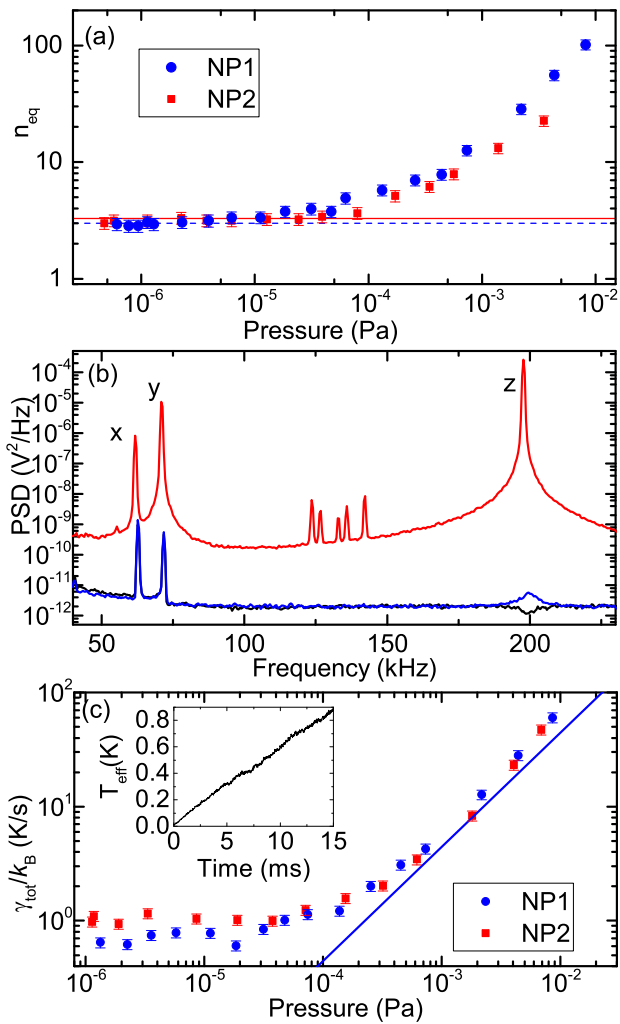}
\caption{(color online). (a) Occupation number along the optical lattice as a function of the pressure for two experimental realizations (Tab.~\ref{tab:npparam}). The gain for feedback cooling is optimized at each pressure. Error bars indicate typical thermal fluctuations in determining $T_{\rm eff}$~\cite{fnote7}. Theoretically expected values of $n_{\rm eq}$ at low pressures, given by Eq.(\ref{eq:teff}), are shown by blue dashed and red solid lines for NP1 and NP2, respectively.  (b) PSDs of the motion of trapped nanoparticles (NP1) with (black and blue curves for the IL and OL signals, respectively, at $\unit[1.3\times 10^{-6}]{Pa}$) and without feedback cooling (red curve, the OL signal at $\unit[9.9]{Pa}$).  (c) Heating rate as a function of the pressure for two experimental realizations. The vertical errors indicate typical thermal fluctuations. A solid line shows calculated values for NP1 with the kinetic theory~\cite{jain2016direct,iwasaki2019electric,tebbenjohanns2019cold}. In the inset, a typical time evolution of $T_{\rm eff}$ in the absence of feedback cooling is shown, where the signal is averaged over 256 times. At low pressures, the heating rate is dominated by the photon recoil heating.}
\label{fig:tvsp}
\end{figure}

In this letter, we show that single charged nanoparticles trapped in an optical lattice are brought to near their ground state along the optical lattice at an ambient temperature via electric feedback cooling, namely, cold damping~\cite{cohadon1999cooling,bushev2006feedback,poggio2007feedback,wilson2015measurement,iwasaki2019electric,tebbenjohanns2019cold,conangla2019optimal}. To make the impact of the laser phase noise (LPN) negligible, we minimize the distance of the trap from the retro-reflecting mirror and besides stabilize the laser frequency to an optical resonator. We establish a model directly connecting the independently measured LPN and the heating rate of nanoparticles' motion and find that the heating rate increases as the LPN increases. Our model shows that the LPN near the oscillation frequency plays a crucial role in the heating dynamics. Furthermore, in our setup, the impact of the LPN is negligible and the residual heating mechanism limiting the lowest attainable occupation number is random photon scattering of the trapping laser. Our results are also useful in other fields involving an optical lattice, such as cold atoms and molecules, for evaluating the impact of the LPN on trapped particles. 


In our experiments, we trap silica nanoparticles with radii of about $R=\unit[220]{nm}$ in a one-dimensional optical lattice formed with a fiber laser [Fig.~\ref{fig:expset}(a)]~\cite{yoneda2017thermal,yoneda2018spontaneous,iwasaki2019electric}. The radii of nanoparticles in the present work are several times larger than typical values in previous studies~\cite{li2011millikelvin,gieseler2012subkelvin,vovrosh2017parametric,tebbenjohanns2019cold,conangla2019optimal,delic2020cooling,tebbenjohanns2020motional}. We load nanoparticles into the optical lattice at around $\unit[600]{Pa}$ by blowing up silica powders with a pulsed laser, similarly to light-induced desorption of nanoparticles attached on a surface~\cite{asenbaum2013cavity,bykov2019direct}. After we load nanoparticles in one of lattice sites, we sweep the laser frequency and adjust their position to the focus of the incident beam where the oscillation frequency along the optical lattice $\Omega_0$ takes a maximum value.

We observe the three-dimensional motion of nanoparticles with two photodetectors measuring the spatio-temporal variation of the infrared light scattered by them~\cite{vovrosh2017parametric}, where one of the detectors is used as an in-loop (IL) detector for feedback cooling and the other is used as an out-of-loop (OL) detector for measuring the motional temperatures $T_{\rm eff}$~\cite{gieseler2012subkelvin,vovrosh2017parametric,tebbenjohanns2019cold}. $T_{\rm eff}$ is obtained by comparing the integrated area of the power spectral densities (PSDs) of nanoparticles with and without cooling~\cite{gieseler2012subkelvin,vovrosh2017parametric}. Using the measured position information, we realize three-dimensional electric feedback cooling of charged nanoparticles by applying three-dimensional electric fields synchronized to their motion such that their motional amplitudes are attenuated in all directions~\cite{iwasaki2019electric}.

\begin{table}
\caption{\label{tab:npparam} Experimental parameters for two realizations. }
\begin{ruledtabular}
\begin{tabular}{cccc}
$$ & ${\rm Power~(mW)}$ & $\Omega_0~{\rm (2 \pi \cdot kHz)}$ & $\hbar\Omega_0/{\rm k_{\rm B}~(\mu K)}$ \\ \hline
${\rm NP1}$ & $168$  & $200$ & $9.6$ \\
${\rm NP2}$ &  $295$  & $256$ & $12$ \\
\end{tabular}
\end{ruledtabular}
\end{table}

When the pressure is decreased to below $\unit[1\times 10^{-5}]{Pa}$, under the condition that the LPN is minimized, the occupation number along the optical lattice, $n_{\rm eq}=T_{\rm eff}/\hbar \Omega_0 -1/2$, is nearly independent from the pressure and approaches the minimum value of about three, suggesting that the influence of background gases becomes negligible at this pressure region~[Fig.\ref{fig:tvsp}(a),(b)]. This fact is further confirmed by measuring the magnitude of the heating via observing the time evolution of the amplitude of the oscillation signal from the OL detector~\cite{iwasaki2019electric,tebbenjohanns2019cold}. The measured heating rate as a function of the pressure [Fig.~\ref{fig:tvsp}(c)] is in agreement with the kinetic theory at high pressures~\cite{gieseler2012subkelvin,vovrosh2017parametric,iwasaki2019electric}, while it is nearly independent from the pressure at pressures lower than $\unit[1\times 10^{-5}]{Pa}$. The heating rate observed in this pressure region is about $\unit[k_{\rm B}\times 1]{K/s}$, where $k_{\rm B}$ is the Boltzmann constant, in agreement with previous studies on photon recoil heating~\cite{jain2016direct}. In what follows, we work at this pressure region to reveal to what extent the LPN and the photon recoils contribute to the observed heating rate for two individual experimental realizations with different laser powers (Tab. ~\ref{tab:npparam}).

\begin{figure}[t]
\includegraphics[width=0.99\columnwidth] {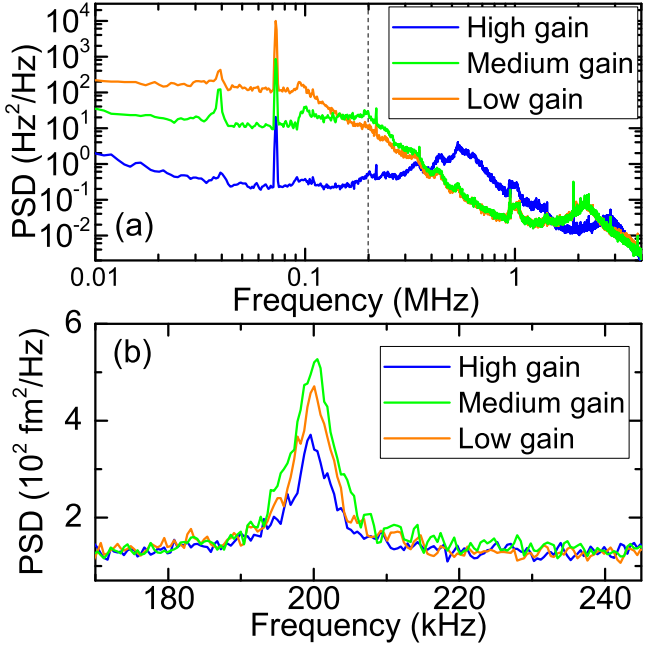}
\caption{ (a) PSDs of the LPN for three values of the feedback gain for stabilizing the laser frequency. With the high gain, the LNP below $\unit[300]{kHz}$ is minimized and we achieve recoil-limited feedback cooling as shown in Fig.~\ref{fig:tvsp}(a).  The oscillation frequency for NP1 is shown by a dashed line. The peaks below $\unit[100]{kHz}$ are due to the bandwidth of the fiber laser itself. (b) PSDs of the OL signal of NP1 for the three conditions of (a). The pressure is $\unit[1.3 \times 10^{-6}]{Pa}$.}
\label{fig:hrvspn}
\end{figure}

The magnitude of the heating via the LPN increases with the distance between the trap position and the retro-reflecting mirror~[Fig.~\ref{fig:expset}(b)]. In our previous setup with the retro-reflecting mirror placed outside the vacuum chamber~\cite{iwasaki2019electric}, we observed that $T_{\rm eff}$ along the optical lattice was limited to a few mK.  We found that the LPN of the fiber laser was the source of heating and improved our setup such that the retro-reflecting mirror is placed at a distance of $d=\unit[14.5]{mm}$ from the trap position, which is more than one order of magnitude shorter than that of typical experiments with optical lattices and our previous work~\cite{iwasaki2019electric,fnote7}. Furthermore, we decrease the LPN itself by several orders of magnitude via stabilizing the frequency of the laser with a high-finesse resonator. In our setup, we are able to tune the magnitude of the LPN by controlling the feedback gain for the frequency stabilization on the feedback circuit. We estimate the PSD of the LPN by multiplying $\alpha^2$ by the PSD of the error signal from the resonator, with $\alpha$ being the slope of the error signal near the resonance. Figure \ref{fig:hrvspn}(a) shows the PSD of the LPN for three values of the feedback gain, where the relative magnitude of the feedback gain is designated by high, medium, and low. Among the three gain values, we observe a dramatic variation in $T_{\rm eff}$ [Fig.~\ref{fig:hrvspn}(b)]. $T_{\rm eff}$ is the lowest at the high gain as expected, while we observe the highest $T_{\rm eff}$ among the three at the medium gain. In Fig.~\ref{fig:hrvspn}(a), we clearly observe that the LPN near the oscillation frequency is the largest at the medium gain. Thus, qualitatively we find that the LPN near the oscillation frequency has a dominant contribution to the observed heating.

To quantitatively analyze our observations, we hereby introduce a theoretical model for the motion of nanoparticles based on previous studies~\cite{jain2016direct,iwasaki2019electric,tebbenjohanns2019cold}. We describe the one-dimensional motion of nanoparticles along the optical lattice in the presence of fluctuating forces and damping mechanisms,

\begin{equation}
\label{eq:eom}
\ddot{q}+\Gamma_{\rm tot}\dot{q}+\Gamma_{\rm c}\dot{q_{\rm n}}+\Omega_0^2(q+P_{\rm n})=\dfrac{F_{\rm BG}+F_{\rm r}}{m}
\end{equation}
with $\Gamma_{\rm tot} = \Gamma_{\rm BG}+\Gamma_{\rm c}+\Gamma_{\rm r}+\Gamma_{\rm p}$. Here $q$ and $q_n$ denote the position of nanoparticles and the noise in the feedback signal, respectively, while $\Gamma_{\rm BG}$, $\Gamma_{\rm c}$, $\Gamma_{\rm r}$, and $\Gamma_{\rm p}$ denote the damping rate due to collisions with background gases, the damping rate due to feedback cooling, the damping rate due to photon recoils, and the damping rate due to the LPN, respectively. In addition, $m$, $P_{\rm n}$, $F_{\rm BG}$, and $F_{\rm r}$ denote the mass of trapped nanoparticles, the stochastic force from the LPN, the stochastic force from background gases, and the stochastic force from photon scattering, respectively~\cite{fnote7}. $T_{\rm eff}$ for the particle following Eq.(\ref{eq:eom})  is given as

\begin{align}
\label{eq:temp}
T_{\rm eff}=&T_0\dfrac{\Gamma_{\rm BG}}{\Gamma_{\rm tot}}+\dfrac{m \Omega_0^2 S_{\rm n}\Gamma_{\rm c}^2}{2k_{\rm B}\Gamma_{\rm tot}}+\dfrac{\hbar \omega_0 P_{\rm sc}}{5m c^2 k_{\rm B}\Gamma_{\rm tot}}+\dfrac{m\Omega_0^6}{k_{\rm B}} I(\Gamma_{\rm tot}) \\ \notag
I(\Gamma_{\rm tot})=&\dfrac{1}{2\pi}\int_{-\infty}^{\infty}\dfrac{\tilde{P_{\rm n}}^2(\omega)d\omega}{(\Omega_0^2-\omega^2)^2+\omega^2\Gamma_{\rm tot}^2}
\end{align}
where $T_0$, $S_{\rm n}$, $\omega_0$, $P_{\rm sc}$, $c$, $\tilde{P_{\rm n}}(\omega)$ are the temperature of background gases, the PSD of $q_{\rm n}$, the frequency of the trapping light, the optical power scattered by nanoparticles, the speed of light, and the Fourier transform of $P_{\rm n}$, respectively. The stochastic forces $F_{\rm BG}$ and $F_{\rm r}$ in Eq.(\ref{eq:eom}) are characterized by $T_0$ and $P_{\rm sc}$, respectively. $\tilde{P_{\rm n}}^2(\omega)$ is obtained by multiplying $\lambda d/c$ by the PSD of the LPN~[Fig.\ref{fig:hrvspn}(a)]. 

In general, the integral of Eq.(\ref{eq:temp}) has to be numerically evaluated. However, considering that the spectral width of the integrand $\Gamma_{\rm tot}$ is much smaller than $\Omega_0$, to a good approximation we can write the integral as 

\begin{align}
\label{eq:pn_int}
 I(\Gamma_{\rm tot})=\dfrac{\tilde{P_{\rm n}}^2(\Omega_0)}{2\Omega_0^2 \Gamma_{\rm tot}}
\end{align}

We confirm that this approximation does not deviate from the numerically obtained values by more than $2\%$ with $\Gamma_{\rm tot} < 2\pi \cdot \unit[300]{Hz}$, which is satisfied for the measurement of $\gamma_{\rm tot}$~\cite{fnote7}. 

\begin{figure}[t]
\includegraphics[width=0.99\columnwidth] {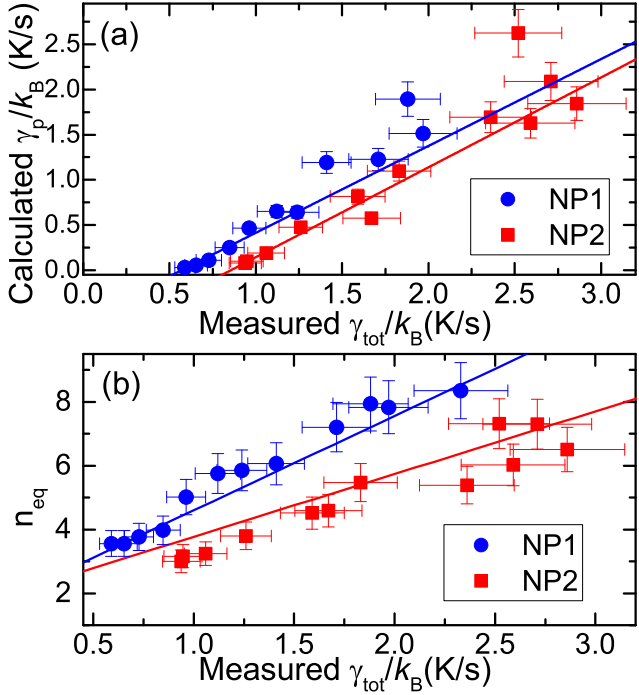}
\caption{ (a) Comparison between the calculated values of $\gamma_{\rm p}$ and the measured values of $\gamma_{\rm tot}$ at $\unit[1.3 \times 10^{-6}]{Pa}$. The solid lines are linear fits on the data points. The horizontal and vertical error bars are due to systematic errors in thermal fluctuations and the error in determining $\alpha$, respectively.  (b) $n_{\rm eq}$ as a function of the measured $\gamma_{\rm tot}$ at $\unit[1.3 \times 10^{-6}]{Pa}$. $n_{\rm eq}$ is measured with varying the feedback gain for stabilizing the laser frequency while $\Gamma_{\rm c}$ is kept constant. The horizontal and vertical error bars indicate typical thermal fluctuations. The solid lines show calculations with Eq.(\ref{eq:teff}) with measured values of $\Gamma_{\rm c}$ and are not fits.}
\label{fig:hrvspn2}
\end{figure}

At low pressures, in the absence of feedback cooling, the heating rate $\gamma_{\rm tot}$ as observed in Fig.~\ref{fig:tvsp}(c) is obtained from $\Gamma_{\rm r}$ and $\Gamma_{\rm p}$ as $\gamma_{\rm tot} = \gamma_{\rm p}+\gamma_{\rm r}$ with $\gamma_{\rm p}=E_{\rm p}\Gamma_{\rm p}$ and $\gamma_{\rm r}=E_{\rm r}\Gamma_{\rm r}$, where $E_{\rm r}$ and $E_{\rm p}$ denote the energies of nanoparticles at equilibrium for photon recoil heating and for LPN heating, respectively.  $E_{\rm r}$ and $E_{\rm p}$ are approximately given by the photon energy $\hbar \omega_0$ and the potential depth, respectively, and are both of the order of $k_{\rm B} \times \unit[10^4]{K}$. Since $\gamma_{\rm tot}$ is directly connected to the equilibrium occupation number as~\cite{jain2016direct},

\begin{align}
\label{eq:neq}
n_{\rm eq} + \dfrac{1}{2}= \dfrac{\gamma_{\rm tot}}{\hbar \Omega_0 \Gamma_{\rm tot}}
\end{align}

we obtain the expression for $\gamma_{\rm tot}$ as follows by comparing Eqs.(\ref{eq:temp},\ref{eq:pn_int},\ref{eq:neq}):

\begin{align}
\label{eq:hr}
\gamma_{\rm tot} = \dfrac{1}{2}m\Omega_0^4 \tilde{P_{\rm n}}^2(\Omega_0)+\dfrac{\hbar\omega_0 P_{\rm sc}}{5mc^2}
\end{align}

The occupation number in the presence of feedback cooling at low pressures $n_{\rm eq}$ is then written as 

\begin{align}
\label{eq:teff}
n_{\rm eq} +\dfrac{1}{2}= \dfrac{1}{2\Gamma_{\rm c}\hbar \Omega_0}\left[ 2\gamma_{\rm tot}+m\Omega_0^2S_{\rm n}\Gamma_{\rm c}^2  \right]
\end{align}
which provides a minimum value of $\sqrt{2\gamma_{\rm tot} m S_{\rm n}}/\hbar$ at an optimum feedback gain.

In Fig.~\ref{fig:hrvspn2}(a), we show a comparison between the measured values of $\gamma_{\rm tot}$ and the calculated values of  $\gamma_{\rm p}$ according to Eq.(\ref{eq:hr}).  For two experimental realizations, we find a good agreement between measurements and calculations. The linear fits on the plots give slopes of 0.96(12) and 0.99(7) for NP1 and NP2, respectively, which are close to the expected value of unity and suggest that the measured increases in $\gamma_{\rm tot}$ are in fact due to the increase in the LPN. The intercepts obtained from the fits are $k_{\rm B} \times 0.51(7)$ and $ k_{\rm B} \times \unit[0.79(7) ]{K/s}$ for NP1 and NP2, respectively. These values are also in agreement with our calculation on  $\gamma_{\rm r}$ providing the values of $k_{\rm B} \times 0.46$ and $ k_{\rm B} \times \unit[0.76 ]{K/s}$ for NP1 and NP2, respectively~\cite{fnote7}. These agreements imply that the approximation of Eq.(\ref{eq:pn_int}) is valid and nanoparticles are heated dominantly by the LPN near their oscillation frequency. Furthermore, we find that $n_{\rm eq}$ observed as a function of the measured values of $\gamma_{\rm tot}$ [Fig.~\ref{fig:hrvspn2}(b)] is in good agreement with Eq.(\ref{eq:teff}). In Fig.~\ref{fig:hrvspn2}(a), we clearly observe that, at a maximum feedback gain for stabilizing the frequency of the laser, $\gamma_{\rm tot}$ approaches the values determined solely by photon recoil heating, indicating that the impact of the LPN is made negligible in our system. Thus, we can safely claim that the lowest $n_{\rm eq}$ obtained in the present study is limited purely by photon recoil heating. 


The impact of the LPN in the presence of a standing wave potential has been of great concern in recent studies on cavity cooling nanoparticles in a single-beam optical trap. A standing wave potential in an optical resonator shaken by the LPN has a significant heating effect~\cite{meyer2019resolved}, where the occupation number is limited to about $2 \times 10^3$. Another recent work in the same direction has shown that nanoparticles are cooled to their ground state when they are placed at the intensity {\it minimum} of the standing wave potential and are supported by another laser trap~\cite{delic2020cooling}, as opposed to our situation where nanoparticles are levitated by an optical lattice alone and stay at the intensity {\it maximum}.  Our study offers a distinct approach for circumventing the strong heating issue associated with a standing wave potential.

For the use of our system in future quantum mechanical experiments, the occupation number of $n_{\rm eq}<1$ is preferred, indicating that $S_{\rm n}$ has to be decreased by a factor of about six for the current values of $\gamma_{\rm r}$. It is feasible to achieve such a sensitivity via improving the optical setup to enhance the efficiency of collecting the light scattered by nanoparticles~\cite{tebbenjohanns2019optimal}. Enhancing the detection sensitivity is important not only for cooling but also for future intriguing physics, where the capability of detecting the zero-point motion of single nanoparticles may allow the direct observation and manipulation of their quantum behaviors.  We also draw an attention that the position sensitivity of $\unit[1.0\times 10^{-28}]{m^2/Hz}$ achieved in the present work [Fig.~\ref{fig:hrvspn}(b)] is orders of magnitude higher than previous studies~\cite{delic2019cavity,tebbenjohanns2020motional} and suggests an application of our system in sensitive accelerometers. 


In conclusion, we demonstrate direct feedback cooling of single nanoparticles in an optical lattice to near their ground state in high vacuum. Both the occupation number and the heating rate along the optical lattice are reduced as the LPN is decreased, in good agreement with calculations based on a model taking into account the effect of the LPN. Under the minimum LPN in our setup, the occupation number is limited purely by photon recoil heating. Our results show that the reduction of the LPN near the oscillation frequency of nanoparticles is crucial for working with them near the ground state. Although photon recoil heating is an unavoidable issue with optical levitation, the occupation number can be further reduced by improving the optical setup. Once cooled to the ground state, our system will provide an important testbed for investigating the quantum superposition states of macroscopic objects~\cite{romero-isart2011large,bassi2013models} with masses of the order of $\unit[0.1]{pg}$. In comparison with recent other approaches~\cite{tebbenjohanns2020motional,delic2020cooling}, our system possesses a qualitatively new control knob for manipulating the position of nanoparticles and for modulating the optical potential via the laser frequency, opening up the possibility to explore the physics with a time-dependent optical lattice~\cite{lignier2007dynamical,hauke2012non,ha2015roton} and anharmonic potentials~\cite{schmelcher2018driven,dhar2019run} at a single-particle level.


\begin{acknowledgments}
We thank M.\,Kozuma for fruitful discussions. We are grateful to T.\,Naruki for his experimental assistance at the early stage of the experiments. This work is supported by the Murata Science Foundation, the Mitsubishi Foundation, the Challenging Research Award, the 'Planting Seeds for Research' program, and STAR Grant funded by the Tokyo Tech Fund, Research Foundation for Opto-Science and Technology, JSPS KAKENHI (Grants No. JP16K13857, JP16H06016, and JP19H01822), and JST PRESTO (Grant No. JPMJPR1661).
\end{acknowledgments}

\section{Supplementary Information}
\subsection{Experimental setup}

A single-frequency infrared laser at a wavelength of $\unit[1550]{nm}$ is focused with an objective lens (NA$=0.85$) and is approximately quarter of the incident power is retro-reflected to form a standing-wave optical trap (an optical lattice). The laser power is $\unit[168]{mW}$ for NP1 and $\unit[295]{mW}$ for NP2, while the mass is nearly equal for both realizations. The retro-reflecting mirror is placed inside the metal housing of the lens such that the distance between the trap position and the mirror is minimized to about $\unit[14.5]{mm}$. The setup around the trap region is installed in a vacuum chamber equipped with a scroll pump, a turbo-molecular pump, an ion pump, and a titanium sublimation pump. 

The trapping light is linearly polarized along the $x$ direction. The light scattered by nanoparticles and the retro-reflected light are both extracted from the isolator and is used for detecting the motion of nanoparticles via two home-made balanced photodetectors, which subtracts the signal without nanoparticles from the signal with nanoparticles. The subtraction is to minimize the influence of the laser intensity noise on the signal. The nanoparticle's motion in the $z$ direction modulates the total intensity of the trapping beam, while those in the $x,y$ directions modulate the spatial beam profile. The light from the trapping beam is incident on the photodetector with a slight misalignment in both $x$ and $y$ directions such that the spatial intensity modulation is also detected. In this way, we observe the three dimensional motion of a trapped nanoparticle with a single photodetector. In the present work, one of the two detectors is used as an in-loop (IL) detector for feedback cooling, while the other detector works as an out-of-loop (OL) detector for measuring the temperature of nanoparticles. 

We load nanoparticles by blowing up silica powders with a pulsed laser at $\unit[532]{nm}$ at pressures of about $\unit[600]{Pa}$. A sample of silica powders is placed about $\unit[1]{mm}$ below the trap region. As compared to the previously well-known method of introducing a mist of ethanol including nanoparticles, our approach does not contaminate the vacuum chamber significantly and makes it easier to evacuate it to high vacuum. Because many lattice cites have deep potentials sufficient for trap nanoparticles, initially nanoparticles are not always trapped at the focal position. We shift their position by sweeping the laser frequency such that they stay at the focal position.

After loading nanoparticles, we evacuate the chamber with a scroll pump and a turbo-molecular pump. At around $\unit[1 \times 10^{-5}]{Pa}$, we open the two valves for separating an ion pump and a titanium-sublimation pump from the main chamber. In this way, we bring nanoparticles from about $\unit[600]{Pa}$ to about $\unit[3\times 10^{-7}]{Pa}$ in several hours.

We generate the feedback signal by sending the IL photodetector signal through high pass filters and low pass filters such that the phase difference of $\unit[90]{^\circ}$ between the photodetector signal and the oscillator output signal is introduced. The feedback signal is applied to the metal housing of the lens pairs. We find that a proper phase control allows us to use the signal for cooling the motion in the $z$ direction also for cooling the motions in the $x$ and $y$ directions to below $\unit[1]{K}$, presumably because of the weak electric fields created by the lens pairs perpendicular to the light propagation direction. Thus, in the present work we do not need electrodes perpendicular to the light propagation direction. As compared to the scheme based on a lock-in amplifier previously we had employed, the present approach provides a higher feedback bandwidth and a better cooling performance.

The frequency of the fiber laser is stabilized to an optical resonator with a finesse of 7800 by means of an acousto-optic modulator (AOM) and an electro-optic phase modulator (EO-PM). The AOM works as a main feedback component covering the frequency range of below $\unit[600]{kHz}$, while the EO-PM has a bandwidth of about $\unit[2.5]{MHz}$ and is required to suppress the oscillation of the AOM at around $\unit[600]{kHz}$. In addition to the phase stabilization, we use an electro-optic amplitude modulator (EO-AM) for suppressing the intensity noise to the level of shot noise.

\subsection{Estimation of the parameters in Eqs.~(1,2)}

In this section, we describe how to obtain parameters in Eqs.~(1,2). 

$\Omega_0$:  We obtain $\Omega_0$ from the PSD of trapped nanoaprticles as shown in Fig.~2(b).

$T_{\rm eff}$: The temperatures of nanoparticles $T_{\rm eff}$ are obtained by comparing the areas of the PSDs with and without cooling, as has been performed in previous studies. To avoid the influence of the increase in the internal temperature of nanoparticles at high vacuum due to laser absorption~\cite{hebestreit2018measuring}, we take the uncooled data at around $\unit[10]{Pa}$. We find that the typical thermal fluctuation of the area of the PSD is lower than $\unit[7]{\%}$ for both cooled and uncooled data. Thus, we estimate the error in determining $T_{\rm eff}$ to be about $\unit[10]{\%}$. 

$\Gamma_{\rm BG}$: After we cool the motion of nanoparticles to $T_{\rm eff}<\unit[10]{K}$, we abruptly turn off electric fields for feedback cooling and observe an exponential growth in $T_{\rm eff}$. The exponential time constant provides $\Gamma_{\rm BG}$. 

$m$: The mass of trapped nanoparticles is estimated by following the kinetic theory relating $\Gamma_{\rm BG}$, the pressure, and the radius of particles~\cite{gieseler2012subkelvin,vovrosh2017parametric,iwasaki2019electric}. For this measurement, we use a capacitance gauge that gives pressure values with an accuracy of $\unit[0.5]{\%}$. The error in estimating the mass of nanoparticles is typically about $\unit[10]{\%}$, which is limited by thermal fluctuations.

$\Gamma_{\rm c}$: When we abruptly turn on the electric fields for feedback cooling, we observe an exponential decrease in $T_{\rm eff}$. The exponential time constant provides $\Gamma_{\rm c}$.

$\Gamma_{\rm r}, \Gamma_{\rm p}$: In principle, $\Gamma_{\rm r}$ and $\Gamma_{\rm p}$ are obtained by $\Gamma_{\rm r}=\gamma_{\rm r}/E_{\rm r}$ and $\Gamma_{\rm p}=\gamma_{\rm p}/E_{\rm p}$, respectively. However, in the present study, the knowledge on $\Gamma_{\rm r}$ and $\Gamma_{\rm p}$ is not required because we can relate experimentally measurable values $\gamma_{\rm r}$, $\gamma_{\rm p}$, and $n_{\rm eq}$ as shown in Eqs.~(4-6).

$S_{\rm n}$: We obtain $S_{\rm n}$ from the PSD of uncooled nanoparticles at around $\unit[10]{Pa}$. We assume $T_{\rm eff}=\unit[300]{K}$ for nanoparticles without feedback cooling and obtain a conversion factor between the photodetector voltage and the position. Then, the noise floor of the detector, determined by the shot noise at photodetectors, provides $S_{\rm n}$.

$P_{\rm sc}$: As detailed in the next section, we estimate $P_{\rm sc}$ using the intensity of the trapping laser.

\subsection{Estimation of the photon scattering rate}

The estimation of the photon recoil heating rate $\gamma_{\rm r}$ requires the values of the photon scattering power $P_{\rm sc}$, which is obtained by multiplying the scattering cross section by the laser intensity. For estimating the laser intensity, the values of the beam waists are crucially important. We determine the beam waists such that the observed oscillation frequencies are consistent with the calculated values. 

In our setup, the optical lattice is formed by an incident elliptic beam and a reflected nearly round beam with a waist larger than that of the incident beam. The oscillation frequencies in this trap are given by

\begin{align}
\label{eq:freq}
\Omega_x=&\sqrt{\dfrac{4A}{m}\left( \sqrt{\dfrac{2P_{\rm i}}{\pi w_x w_y}}+ \sqrt{\dfrac{2P_{\rm o}}{\pi w_0^2}}\right)} \\ \notag
&\times \sqrt{\left( \dfrac{1}{w_x^2} \sqrt{\dfrac{2P_{\rm i}}{\pi w_x w_y}} + \dfrac{1}{w_0^2} \sqrt{\dfrac{2P_{\rm o}}{\pi w_0^2}}\right)} \\
\Omega_y=&\sqrt{\dfrac{4A}{m}\left( \sqrt{\dfrac{2P_{\rm i}}{\pi w_x w_y}}+ \sqrt{\dfrac{2P_{\rm o}}{\pi w_0^2}}\right)} \\ \notag
&\times \sqrt{\left( \dfrac{1}{w_y^2} \sqrt{\dfrac{2P_{\rm i}}{\pi w_x w_y}} + \dfrac{1}{w_0^2} \sqrt{\dfrac{2P_{\rm o}}{\pi w_0^2}}\right)} \\
\Omega_z =&\dfrac{8\pi}{\lambda}\sqrt{\dfrac{A}{\pi m}\sqrt{\dfrac{P_{\rm i}P_{\rm o}}{w_x w_y w_0^2}}} \\
A =& \dfrac{2\pi R^3}{c} \dfrac{n^2-1}{n^2+2}
\end{align}
where $P_{\rm i}$, $P_{\rm o}$, $w_x$, $w_y$, $w_0$, and $n$ are the power of the incident beam, the power of the reflected beam, the beam waist of the incident beam in the $x$ direction, the beam waist of the incident beam in the $y$ direction, the beam waist of the reflected beam, and the refractive index of nanoparticles, respectively. We use $n \simeq 1.45$ for the refractive index at $\lambda = \unit[1550]{nm}$.  From the observed oscillation frequencies at a given power, we deduce $w_x \simeq 1.25 \mu m$, $w_y\simeq 1.15\mu m$, and $w_0 \simeq 2 \mu m$. The power scattered by trapped nanoparticles is then calculated as~\cite{jain2016direct} 

\begin{align}
P_{\rm sc}=&\dfrac{32\pi^3 A^2c^2}{3\lambda^4} I_0 \\
I_0 =& \left( \sqrt{\dfrac{2P_{\rm i}}{\pi w_x w_y}}+ \sqrt{\dfrac{2P_{\rm o}}{\pi w_0^2}} \right)^2
\end{align}

When the power of the incident beam is $\unit[300]{mW}$, calculated values are $I_0 \simeq \unit[20]{MW/cm^2}$ and $P_{\rm sc} \simeq \unit[3.7]{mW}$ for $R\simeq \unit[220]{nm}$.

\subsection{Validity of the approximation for estimating the influence of the phase noise}

\begin{figure}[t]
\includegraphics[width=0.6\columnwidth] {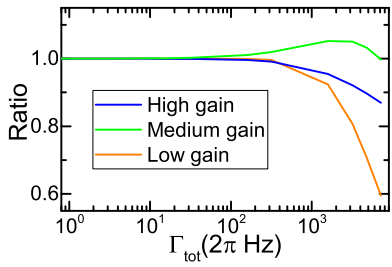}
\caption{The ratio of the approximated values in Eq.(3) to the numerically obtained values for three values of the feedback gain for stabilizing the laser. The data are the same as those in Fig.3(a). }
\label{fig:hrvspn3}
\end{figure}

We numerically confirm that the approximation of the integral in Eq.(3) is valid when $\Gamma_{\rm tot}$ is sufficiently small. This condition applies to the measurement of $\gamma_{\rm tot}$. Figure~\ref{fig:hrvspn3} shows the ratio of the approximated values to the numerically obtained values. For large values of $\Gamma_{\rm tot}$, the approximation provides lower values than numerical values. In the presence of feedback cooling, we have a typical value of $\Gamma_{\rm c}\approx \unit[2\pi \cdot 6]{kHz}$, resulting in a deviation of more than a few $\%$ between the approximated values and the numerical values. 

\bibliographystyle{apsrev}


\end{document}